\documentclass[hyper]{JHEP} 

\usepackage{epsfig}

\newcommand\fverb{\setbox\pippobox=\hbox\bgroup\verb}

\newcommand\fverbdo{\egroup\medskip\noindent%

            \fbox{\unhbox\pippobox}\ }

\newcommand\fverbit{\egroup\item[\fbox{\unhbox\pippobox}]}

\newbox\pippobox

\title{Ho\v{r}ava-Lifshitz Gravity And Ghost Condensation}
\author{J. Kluso\v{n}\\
Department of
Theoretical Physics and Astrophysics\\
Faculty of Science, Masaryk University\\
Kotl\'{a}\v{r}sk\'{a} 2, 611 37, Brno\\
Czech Republic\\
E-mail: \email{klu@physics.muni.cz}}
\preprint{\hepth{1008.5297}}

 \abstract{In this paper we formulate
 RFDiff invariant $f(R)$  Ho\v{r}ava-Lifshitz gravity
 and we show that it is related to the ghost condensation
 in the projectable version of Ho\v{r}ava-Lifshitz gravity.}
\keywords{Ho\v{r}ava-Lifshitz gravity, Ghost Condensation}

\def\hc{\hat{c}}
\def\mH{\mathcal{H}}

\def\iD{\left(D^{-1}\right)}

\def\bx{\mathbf{x}}
\def\by{\mathbf{y}}

\newcommand{\hG}{\hat{G}}

\newcommand{\mG}{\mathcal{G}}

\newcommand{\hK}{\hat{K}}

\newcommand{\bT}{\mathbf{T}}

\newcommand{\mL}{\mathcal{L}}

\def\pb #1{\left\{#1\right\}}

\begin{document}
\section{Introduction and Summary}\label{first}
Recently Pet\v{r} Ho\v{r}ava formulated idea
considering consistent renormalizable quantum theory of gravity
\cite{Horava:2009uw} (see also
\cite{Horava:2008jf,Horava:2008ih,Horava:2009if,Horava:2010zj}). This
proposal is based on an idea that the ultra-violet (UV)
behavior of quantum gravity is improved thanks to
terms with higher spatial derivatives where at the
same time the number of time derivatives in the Lagrangian
remains equal to two so that there is no problem with ghosts
that arise in  Lorentz invariant higher-derivative theories of gravity.
It is clear that the breaking the symmetry between
space and time we lose the Lorentz Invariance of given
theory that now is not the fundamental symmetry of the theory
and can emerge at low energies as an approximative symmetry.

However it turned out that
even if the Ho\v{r}ava's formulation is very promising and
interesting there are many conceptional problems that arise
in this theory as
in any theory of gravity with reduced diffeomorphism group.
Explicitly, the fact that the theory is not invariant under
full diffeomorphism group implies an existence of new
degrees of freedom which is the mode of helicity zero.
This mode has natural geometrical origin since the introduction
 of the preferred time coordinate leads to the
 foliation of the space-time manifold by  space-like
surfaces where the  helicity-0 mode is  excitation of
this foliation structure. Very interesting analysis of the properties
of  this
mode has been done recently in \cite{Blas:2010hb}. The main
conclusion derived here is that  this extra
 mode does not decouple at low
energies and hence it is questionable whether Ho\v{r}ava-Lifshitz
gravity can flow to General Relativity at low energies.
On the other hand authors in \cite{Blas:2010hb} suggested
very interesting possibility that Ho\v{r}ava-Lifshitz
gravity could flow at low energies to Lorentz violating
model of modified gravity where the modifications
 are small so that they do not contradict to experimental data.
On the other hand the analysis  of the modified gravity
models performed in the past
 shown that properties of these extra modes
imply that these modified gravity modes are not
phenomenologically acceptable
\footnote{For review, see \cite{Blas:2008uz,Bebronne:2009iy,Rubakov:2008nh}.}.
Unfortunately this situation occurs in the original
version of Ho\v{r}ava's proposal where the extra mode
possesses pathological behavior
\cite{Blas:2009yd,Koyama:2009hc,Cerioni:2010uz}.

In \cite{Blas:2010hb} three different models of
Ho\v{r}ava-Lifshitz gravities were studied.
The first one is the  'projectable' version of
 the original proposal \cite{Horava:2009uw} where
the lapse function depends on time only. The
second one is the model with smaller symmetry
group (Denoted as RFDiff invariant theory
in \cite{Blas:2010hb})  and can be considered as a power counting
renormalizable version of the ghost condensation
\cite{ArkaniHamed:2003uy}. However as was nicely
shown in \cite{Blas:2010hb} the  spectrum
of this model possesses the second helicity-0 mode
that unfortunately leads to fast
instabilities or to the break-down of the perturbative description.
These facts imply that it is unclear whether this
model can be a promising candidate for description of
quantum gravity. Finally the third model studied in
\cite{Blas:2010hb} is the so called "healthy-extended"
Ho\v{r}ava-Lifshitz gravity introduced in
\cite{Blas:2009ck,Blas:2009qj}
\footnote{For Hamiltonian analysis of this model,
see \cite{Kluson:2010xx,Kluson:2010nf}.}. It was shown there
 that now the
 scalar sector  does not suffer from pathologies
 and that this model is compatible with
 phenomenological constraints for suitable choices
of parameters. Hence this model can be considered
as a starting point for constructing a
renormalizable theory of quantum gravity.

In this paper we  consider
some aspects of RFDiff invariant Ho\v{r}ava-Lifshitz
theories. Namely we present
construction of these models based on the original
Ho\v{r}ava's "detailed balance condition"
\cite{Horava:2009uw} that was introduced in
\cite{Kluson:2009rk,Kluson:2009xx}. This construction
leads to  models that
are invariant under transformations
\begin{equation}\label{symgr}
t'=t+\delta t \ , \quad x'^i=x^i+\xi^i(\bx) \ ,
\end{equation}
where $\delta t=\mathrm{const}$ and where
$\xi^i$ are space dependent parameters of
spatial diffeomorphism. Then we extend
symmetries of given model when we demand that
the action should be invariant under transformation
\begin{equation}\label{tdif1}
t'= t'=t+\delta t \ , \quad x'^i=x^i+\xi^i(\bx,t) \ .
\end{equation}
that is exactly the symmetry group of RFDiff  invariant
theories. It is important that this symmetry
group is
different from
\emph{foliation preserving diffeomorphism}
that has the form
\begin{equation}
t'=f(t) \ , \quad x'^i=x^i+\xi^i(\bx,t) \ .
\end{equation}
In order to construct action that is
invariant under  (\ref{tdif1})  we have to introduce
the fields $N^i$ that are well  known "shifts"
from $3+1$ analysis of General Relativity. Note
that we do not need to introduce the lapse
function $N$. As a consequence of this fact
the Hamiltonian constraint is absent in theories
invariant under (\ref{tdif1}). In other
words the Hamiltonian of RFDiff invariant
Ho\v{r}ava-Lifshitz theory has the structure of
the Hamiltonian of diffeomorphism invariant theory
where the temporal diffeomorphism has been fixed.
In fact, we explicitly construct such a theory
and we argue that RFDiff invariant Ho\v{r}ava-Lifshitz
theory can be interpreted as a result of the
ghost condensation in Ho\v{r}ava-Lifshitz gravity
coupled to specific form of the scalar field action.
We also argue that in order to have the first
class constraint that can be fixed we should
consider projectable version of Ho\v{r}ava-Lifshitz
gravity.  In fact, it was shown in
  \cite{Henneaux:2009zb} that the Hamiltonian
  constraint in non-projectable Ho\v{r}ava-Lifshitz
  gravity is the second class constraint and
  it certainly does not make sense to fix
  it. On the other hand the projectable
  version of Ho\v{r}ava-Lifshitz gravity is
characterized by the global form of the Hamiltonian
constraint that is trivially the first class
constraint and hence the gauge fixing can be performed.

Let us outline our results. We construct general
form of RFDiff invariant $f(R)$ Ho\v{r}ava-Lifshitz
gravity theories that are generalizations of the
theories studied in \cite{Blas:2010hb}.
 We show that they can be derived
by ghost condensation from the projectable version
of $f(R)$ Ho\v{r}ava-Lifshitz gravities.

The structure of this paper is as follows. In the
next section (\ref{second}) we define
$f(R)$ models of gravity that obey the detailed balance
condition and that are invariant under
symmetry group (\ref{symgr}).
 Then in section (\ref{third}) we extend
 symmetries of given theories so that they
 are invariant under (\ref{tdif1}) and hence
they are RFDiff-invariant $f(R)$ Ho\v{r}ava-Lifshitz
gravities. Finally in section (\ref{fourth}) we
show that these theories can be interpreted
as the ghost condensation in the projectable version
of Ho\v{r}ava-Lifshitz gravity.

\section{$f(R)$ Gravity at Criticality}\label{second}
In this section we introduce
models of  gravity based on
original Ho\v{r}ava's proposal
\cite{Horava:2008ih} and its
generalization for the
the construction of
$f(R)$ Ho\v{r}ava-Lifshitz gravities
performed in
\cite{Kluson:2009rk,Kluson:2009xx}
that are invariant under (\ref{symgr})
\footnote{$f(R)$ Ho\v{r}ava-Lifshitz gravities
were also extensively studied in
\cite{Chaichian:2010yi,Carloni:2010nx,Chaichian:2010zn,Chaichian:2010zn}.}.
Following \cite{Horava:2008ih}
we assume an existence
of  $D+1$ dimensional quantum
 theory of gravity that is characterized
 by following quantum
Hamiltonian density
\begin{eqnarray}\label{QGRH}
\hat{\mH}=\kappa^2 \sqrt{\hat{g}}
\left(\sum_{n=0}^\infty \hc_n(\hat{g}_{ij})
\left(\hat{Q}^{\dag ij} \frac{1}{\hat{g}}
\hat{\mG}_{ijkl}\hat{Q}^{kl}\right)^n-
\hc_0(\hat{g}_{ij})\right) \ ,  \nonumber \\
\end{eqnarray}
where
\begin{equation}\label{hQij}
\hat{Q}^{\dag ij}= -i\hat{\pi}^{ij}
+\sqrt{\hat{g}}\hat{E}^{ij}(\hat{g}_{ij})
\ , \quad \hat{Q}^{
ij}=i\hat{\pi}^{ij}
+\sqrt{\hat{g}}\hat{E}^{ij}(\hat{g}_{ij})
\ ,
\end{equation}
and where $\hat{g}=\det \hat{g}_{ij}$
and $\kappa$ is a coupling constant of
given theory. Note that the fundamental
operators of quantum theory of gravity
are  metric components
 $\hat{g}_{ij}(\bx) \ ,
i=1,\dots,D, \bx=(x^1,\dots,x^D)$ together with their conjugate
momenta $\hat{\pi}^{ij}(\bx)$. These
operators  obey
the commutation relations
\begin{equation}
[\hat{g}_{ij}(\bx),\hat{\pi}^{kl}(\by)]=
\frac{1}{2}(\delta_i^k\delta_j^l+\delta_i^l\delta_j^k)
\delta(\bx-\by) \ .
\end{equation}
Further, $\hat{c}_n$ defined in
(\ref{QGRH}) are scalar functions that
depend on
$\hat{g}_{ij}$ only.
 It is
clear that in the Schr\"{o}dinger
representation the operators
(\ref{hQij}) take the form
\begin{equation}
\hat{Q}^{ij}(\bx)=-\frac{\delta}
{\delta g^{ij}(\bx)}+\sqrt{g}(\bx)E^{ij}(\bx) \ ,
\quad
\hat{Q}^{\dag ij}(\bx)=
\frac{\delta }{\delta  g^{ij}(\bx)}+
\sqrt{g}(\bx)E^{ij}(\bx) \ .
\end{equation}
The next goal is to specify the form of
the operators $E^{ij}$. To do this we
assume that the theory obeys the
\emph{detailed balance condition} so
that
\begin{equation}\label{defE}
\sqrt{g}(\bx)E^{ij}(\bx)=\frac{1}{2}\frac{
\delta W}{\delta g^{ij}(\bx)} \ ,
\end{equation}
where $W$ is an action of $D$
dimensional gravity. As in
\cite{Horava:2008ih}
 we construct the vacuum wave functional of
$D+1$ dimensional theory as
\begin{equation}\label{vvf}
\Psi[g(\bx)]=
\exp\left(-\frac{1}{2}W\right) \ ,
\end{equation}
where $W$ is the Einstein-Hilbert
action in $D$ dimensions
\begin{equation}
W=\frac{1}{2\kappa^2_W}
\int d^D \bx \sqrt{g}R \ .
\end{equation}
Generally the action $W$ could also
contains additional terms that are
functions of metric however the
explicit form of $W$ will not be
important in following discussion.

The form of the vacuum wave functional
(\ref{vvf}) implies that it is
annihilated by
 (\ref{QGRH}). Further as a consequence
of the detailed balance condition  the norm of the
 functional (\ref{vvf})  coincides with the partition
 function of $D$ dimensional Euclidean gravity.
 In other words we have again infinite
 number of possible Hamiltonians that
annihilate the vacuum state (\ref{vvf})
and that are defined using the
principle of detailed balance.

In order to find the Lagrangian
formulation of this  theory we now
consider the classical form of the
Hamiltonian density (\ref{QGRH})
that we can now write in the form
\begin{equation}\label{defmH0}
\mH(t,\bx)=
\kappa^2 \sqrt{g}f\left( Q^{\dag ij}
\frac{1}{g}\mG_{ijkl} Q^{kl}\right)
 \ ,
\end{equation}
where $f$ is an arbitrary function that
can be defined by its Taylor expansion as in
(\ref{QGRH}).
  Further, the functions
 $Q^{ij}$ and
$Q^{\dag ij}$ are defined as
\begin{equation}
Q^{ ij}=i\pi^{ij}+ \sqrt{g}E^{ij} \ ,
\quad Q^{\dag ij}=-i\pi^{ij}+
\sqrt{g}E^{ij} \ , \quad
\end{equation}
where $g_{ij},\ , i,j=1,\dots,D$ are
components of metric and  $\pi^{ij}$
are conjugate momenta. These canonical
variables have non-zero   Poisson
brackets
\begin{equation}
\pb{g_{ij}(\bx),\pi^{kl}(\by)}=
\frac{1}{2}(\delta_i^k\delta_j^l+
\delta_i^l\delta_j^k)\delta(\bx-\by) \ .
\end{equation}
Finally  $\mG_{ijkl}$ denotes the inverse
of the De Witt metric
\begin{equation}\label{DeWi}
\mG_{ijkl}=\frac{1}{2}(g_{ik}g_{jl}+
g_{il}g_{jk})-\tilde{\lambda}g_{ij}g_{kl}
\
\end{equation}
with
$\tilde{\lambda}=\frac{\lambda}{D\lambda-1}$.
The "metric on the space of
metric", $\mathcal{G}^{ijkl}$ is
defined as
\begin{equation}\label{DeW}
\mG^{ijkl}=\frac{1}{2}(g^{ik}g^{jl}+g^{il}
g^{jk}-\lambda g^{ij}g^{kl}) \
\end{equation}
with $\lambda$ an arbitrary real
constant. Note that (\ref{DeWi})
together with (\ref{DeW}) obey the
relation
\footnote{
Note that we use the terminology
introduced in \cite{Horava:2008ih}
and that we review there.
 In case of relativistic
theory, the full diffeomorphism
invariance fixes the value of $\lambda$
uniquely to equal $\lambda=1$. In this
case the object $\mG_{ijkl}$ is known
as the "De Witt metric". We use this
terminology to more general case when
$\lambda$ is not necessarily equal to
one.}
\begin{equation}
\mG_{ijmn}\mG^{mnkl}=\frac{1}{2}
(\delta_i^k\delta_j^l+\delta_i^l\delta_j^k)
\ .
\end{equation}
Using (\ref{defmH0}) we now determine corresponding
Lagrangian. We begin with the canonical
 equation of motion
for $g_{ij}$
\begin{equation}
\partial_t g_{ij}=
\pb{g_{ij},H}=2\kappa^2 \frac{1}{\sqrt{g}}\mG_{ijkl}\pi^{kl}f'
\left( Q^{\dag ij}
\frac{1}{g}\mG_{ijkl} Q^{kl}\right) \ ,
\end{equation}
where $H=\int d^D\bx \mH$ with $\mH$ given
in (\ref{defmH0}) and where $f'(x)=\frac{df}{dx}$.
Using this equation we find
\begin{equation}\label{parttg}
\partial_t g_{ij}\mG^{ijkl}\partial_t g_{kl}=
4\kappa^4  \left(\pi^{ij}\frac{1}{g}\mG_{ijkl}\pi^{kl}\right)
f'^2
\left(\pi^{ij}\frac{1}{g}\mG_{ijkl}\pi^{kl} +
E^{ij}\mG_{ijkl}E^{kl}\right) \ .
 \end{equation}
For further purposes we introduce the notation
\begin{equation}
G=\frac{1}{4\kappa^4}
\partial_t g_{ij}\mG^{ijkl}\partial_t g_{kl} \ ,
\quad
V(g)=
E^{ij}\mG_{ijkl}E^{kl}\ , \quad P=
\pi^{ij}\frac{1}{g}\mG_{ijkl}\pi^{kl} \ .
\end{equation}
At this place we would like to stress that
generally we can abandon the detailed balance
condition and consider $V(g)$ as general
potential for the metric.
Then we can presume that (\ref{parttg})
 can be solved
for $P$ as
\begin{equation}
P=
\Psi\left(G,V\right) \
\end{equation}
so that  $\Psi$ obeys the equation
\begin{equation}
G=\Psi f'^2(\Psi+V) \ .
\end{equation}
Taking derivative of  this equation with respect to $G$
we find the useful relation
\begin{equation}\label{difG}
1=\frac{d\Psi}{dG}f'^2+
2\Psi f'f''\frac{d\Psi}{dG} \ .
\end{equation}
Then it is easy to see that  corresponding Lagrangian takes the form
\begin{eqnarray}\label{Ld}
L&=&\int d^D\bx \mL=\int d^D\bx(\pi^{ij}\partial_t g_{ij}-\mH)=
\nonumber \\
&=&\kappa^2 \int
d^D\bx \sqrt{g}(2\Psi(G,V) f'(\Psi(G,V))-f(\Psi(G,V))) \ .
\nonumber \\
\end{eqnarray}
As the next step we
show that the  action
\begin{equation}\label{SHGG}
S=\int dt  d^D\bx \mL \ ,
\end{equation}
 where  $\mL$ is  given in
(\ref{Ld}) is invariant under the
transformation
\begin{equation}\label{spd}
t'=t+\delta t
\ , \delta t=\mathrm{const} \ , \quad
x'^i=x^i(\bx) \ .
\end{equation}
This follows from the fact that we
presumed that
 the functional $W$ is
invariant under the spatial
diffeomorphism under which the metric
$g_{ij}$
 and tensor $E^{ij}$ transform as
\begin{eqnarray}\label{mettr}
g'_{ij}(\bx')&=&g_{kl}(\bx)\iD^k_i
\iD^l_j  \ , \nonumber \\
E'^{ij}(\bx')&=&
E^{kl}(\bx)D_k^i D_l^j \ ,\nonumber \\
\end{eqnarray}
where
\begin{equation}
 D^i_j=\frac{\partial
x'^i}{\partial x^j} \ , \quad D^i_j
\iD^j_k=\delta^i_k \ .
\end{equation}
Using the transformation property of
$g_{ij}$ we find that the metric
 $\mG_{ijkl}$ transforms
as
\begin{eqnarray}
\mG'_{ijkl}(\bx')=
\mG_{i'j'k'l'}(\bx)\iD^{i'}_i
\iD^{j'}_j \iD^{k'}_k\iD^{l'}_l \ .
\nonumber \\
\end{eqnarray}
Finally, using the fact that $d^D\bx'\sqrt{g'(\bx')}=
d^D\bx \sqrt{g(\bx)}$ we see that
 the invariance of the action under
the  spatial diffeomorphism (\ref{spd})
is obvious.
\section{Extension of Symmetries}\label{third}
We  argued that the Lagrangian
(\ref{Ld}) is invariant under
 $D$-dimensional \emph{time independent spatial
diffeomorphism}.  In
\cite{Horava:2009uw,Horava:2008ih}
 these symmetries were extended to the
diffeomorphism that respect the
preferred codimension-one foliation
$\mathcal{F}$ of the theory by the
slices of fixed time. On the other
hand we make following extension
of symmetries
\begin{equation}\label{tdif3}
t'=t+\delta t \ , \quad  \delta t=\mathrm{const}  \ , \quad
x'^i=x^i+\xi^i(\bx,t) \
\end{equation}
that is RFDiffs symmetry group in terminology of
\cite{Blas:2010hb}. Let us now study  consequences
of the requirement of the invariance of
the action under  (\ref{tdif3}).

We firstly note that the
 metric components $g_{ij}$  transform under
 (\ref{tdif3})  as
\begin{eqnarray}
g'_{ij}(\bx',t')=
g_{ij}(\bx,t)-\partial_i \xi^k(t,\bx)
g_{kj}(\bx,t)-g_{ik}(t,\bx)\partial_j
\xi^k(\bx,t) \ , \nonumber \\
g'^{ij}(\bx',t')=
g^{ij}(\bx,t)+ \partial_k  \xi^i(t,\bx)
g^{kj}(\bx,t)+g^{ik}(t,\bx)\partial_k
\xi^j(\bx,t) \ . \nonumber \\
\end{eqnarray}
However now due to the fact that the
gauge parameter $\xi^i$ depends on time
we find that $\partial_t g_{ij}$ does
not transform covariantly under
(\ref{tdif3}). In order to
find an action that is invariant under
(\ref{tdif3}) it is necessary to
introduce new fields $N_i(t,\bx)$
that transform under (\ref{tdif3}) as
\begin{eqnarray}
N'_i(\bx',t' )&=&
N_i(\bx,t )-g_{ij}(\bx,t)\partial_t \xi^j(\bx,t)
-\partial_i \xi^j(\bx,t) N_j(\bx,t) \ ,
\nonumber \\
N'^i(\bx',t')&=&
N^i(\bx,t)-\partial_t \xi^i (t,\bx)+
N^j(t,\bx)\partial_j \xi^i(t,\bx) \ .
\nonumber \\
 \end{eqnarray}
Let us define
\begin{equation}
\hK_{ij}=\partial_t g_{ij}
-D_i N_j-D_j N_i \ ,
\end{equation}
where $D_i$ is a covariant derivative constructed
from $g_{ij}$ that obeys $D_k g_{ij}=0$.  Then it is easy to see that
\begin{eqnarray}
\hK'_{ij}(\bx',t')
=\hK_{ij}(t,\bx)-\partial_i \xi^k(t,\bx)
 \hK_{kj}(t,\bx)
-\hK_{ik}(t,\bx)\partial_j \xi^k
(t,\bx)\nonumber \\
\end{eqnarray}
that means that $\hK_{ij}$ transforms
covariantly under (\ref{tdif3}). Hence
the natural  generalization of (\ref{SHGG})
takes the form
\begin{equation}\label{Smass}
S=\int dt d^D\bx \mL \ ,
\quad
\mL=\kappa^2 \sqrt{g}(2\Psi(\hat{G},V) f'(\Psi(\hat{G},V))
-f(\Psi(\hat{G},V))) \  ,
\nonumber \\
\end{equation}
where
\begin{equation}
\hat{G}\equiv\frac{1}{4\kappa^2}\hK_{ij}\mG^{ijkl}\hK_{kl} \ .
\end{equation}
This Lagrangian can be considered as
generalization of the
RFDiff-invariant
theories studied in
\cite{Blas:2010hb}.

It is instructive to determine
Hamiltonian from the action (\ref{Smass}).
We firstly  find  canonical
momenta from (\ref{Smass})
\begin{eqnarray}
\pi^{ij}&=&\frac{\delta S}{\delta
\partial_t g_{ij}}=\frac{1}{2\kappa^2}
\sqrt{g}\mG^{ijkl}\pi_{kl}
\left(2\frac{d\Psi}{d\hG}
f'+2\Psi \frac{d\Psi}{d\hG} f''-
f'\frac{d\Psi}{d\hG}\right)=
\nonumber \\
&=&\frac{1}{2\kappa^2}
\sqrt{g}\mG^{ijkl}\hK_{kl}\frac{1}{f'}
 \ , \quad
\pi^i=\frac{\delta S}{\delta\partial_t N_i}
\approx 0 \ ,
\nonumber \\
\end{eqnarray}
where we used (\ref{difG}). Using this
relation we can easily  find corresponding
Hamiltonian
\begin{eqnarray}\label{Hrfd}
H&=&\int d^D \bx(\partial_t g_{ij}\pi^{ij}-L)
=\nonumber \\
&=&\int d^D \bx (
2\kappa^2 \sqrt{g}f+N^i\mH_i) \ ,
\nonumber \\
\end{eqnarray}
where
\begin{equation}
\mH_i=-g_{ik}D_j\pi^{jk} \approx 0
\end{equation}
is standard secondary constraint
related to the primary constraint $\pi^i\approx 0$.
In other words this theory is invariant under
spatial diffeomorphism generated by
\begin{equation}
\bT_S(N^i)=\int d^D\bx N^i \mH_i \ .
\end{equation}
On other hand we see from  the structure of the Hamiltonian
(\ref{Hrfd}) that the Hamiltonian constraint
is absent. In the next section we show that
this Hamiltonian is related to specific form
of ghost condensation.
\section{RFDiff invariant Ho\v{r}ava-Lifshitz Gravity and
Ghost Condensation}\label{fourth}
We again begin with the Hamiltonian formulation of the
 Ho\v{r}ava-Lifshitz gravity
where the Hamiltonian density (without the first
class constraints that generate the spatial diffeomorphism)
 takes the form
\begin{equation}\label{mHTmass}
\mH=\kappa^2 \sqrt{g}
f\left(\pi^{ij}\frac{1}{g}
\mG_{ijkl}\pi^{kl}+E^{ij}\mG_{ijkl}E^{kl}\right) \ .
\end{equation}
Now we presume  that this Hamiltonian density
arises in the process of the specific form of the
 gauge fixing.
Explicitly we  consider the system of the
gravity coupled with scalar field. The dynamics
of this system is governed by Hamiltonian that
is the sum of the first class constraints
\begin{equation}
H^G=\int d^D \bx \mH_0(\bx)+\int d^D\bx N^i\mH_i(\bx) \ ,
\quad \mH^i=-2g_{ik}\nabla_j\pi^{jk}+p_\phi \partial_i\phi \ .
\end{equation}
The canonical variables for the scalar field
are $\phi$ and the momentum conjugate
 $p_\phi$
with non-zero Poisson brackets
\begin{equation}
\pb{\phi(\bx),p_\phi(\by)}=\delta(\bx-\by) \ .
\end{equation}
By presumption $H_0=\int d^D\bx N\mH_0(\bx)\approx 0$ is the first
class constraint so that the Hamiltonian $H^G$
weakly vanishes. We further presume that the
gauge freedom  generated by
 $H_0$ can be fixed by the
gauge fixing condition
\begin{equation}
\mG=\phi(\bx)-t=0 \ .
\end{equation}
In other words, we presume that $\mG$ together
with $H_0$ are the second class constraints
that vanish strongly. As a result we find
that the action of the  gauge fixed theory takes
the form
\begin{equation}\label{gaugefixedaction}
S=\int dt d^D\bx (\pi^{ij}\partial_t g_{ij}+
p_\phi \partial_t \phi-H^G)=
\int dt d^D \bx (\pi^{ij}\partial_t g_{ij}+p_\phi
-N^i\mH_i) \ .
\end{equation}
From (\ref{gaugefixedaction}) we see that
it is natural to identify the Hamiltonian
of the gauge fixed theory with $-p_\phi$.
On the other hand since we know that the Hamiltonian
density of the gauge fixed theory is $\mH$
we have following identification
\begin{equation}
p_\phi=-\mH \
\end{equation}
or equivalently, using (\ref{mHTmass}) we can
rewrite this relation into the form
\begin{equation}
\frac{1}{\kappa^4 g}p^2_\phi=
f^2\left(\pi^{ij}\frac{1}{g}
\mG_{ijkl}\pi^{kl}+E^{ij}\mG_{ijkl}E^{kl}\right)
\ .
\end{equation}
Now we presume that $f^2$
 has an inverse function  that we denote
as $\Psi$. Then we  find
\begin{eqnarray}
-\Psi\left(\frac{1}{\kappa^4 g}p^2_\phi\right)+
\pi^{ij}\frac{1}{g}
\mG_{ijkl}\pi^{kl}+E^{ij}\mG_{ijkl}E^{kl}=0 \ .
\end{eqnarray}
This equation can be interpreted as
the strongly vanishing constraint $\mH_0$
\begin{equation}\label{mH0}
\mH_0=-\kappa^2 \sqrt{g}\Psi\left(\frac{1}{\kappa^4 g}p^2_\phi\right)
+\kappa^2 \left(\pi^{ij}\frac{1}{\sqrt{g}}
\mG^{ijkl}\pi_{kl}+\sqrt{g}E^{ij}\mG_{ijkl}E^{kl}\right)
=0 \ .
\end{equation}
Clearly the Poisson bracket between $\mH_0$ and
$\mG$ is non-zero that confirms that the constraints
$\mH_0$ together with $\mG$ are the second class constraints.

Knowing the form of the Hamiltonian constraint
$\mH_0$ we can find the Lagrangian density for the
given system. As the first step we
find the relation between $\partial_t \phi$ and
canonical variables
\begin{eqnarray}\label{partialtphi}
\partial_t \phi=\pb{\phi,H^G}=
-2N\frac{1}{\kappa^2 \sqrt{g}}p_\phi\Psi'
 \ .
\nonumber \\
\end{eqnarray}
The equation  (\ref{partialtphi}) implies
\begin{equation}\label{parttphi}
\frac{1}{4N^2}(\partial_t\phi)^2=
\frac{p^2_\phi}{\kappa^4 g}\Psi'^2\left(
\frac{p^2_\phi}{\kappa^4 g}\right) \ .
\end{equation}
Now we presume that this equation can be solved
for $\frac{1}{\kappa^4 g}p^2_\phi$ as
\begin{equation}
\frac{1}{\kappa^4 g}p^2_\phi=\Phi\left(\frac{1}{4N^2}
(\partial_t\phi)^2\right) \ .
\end{equation}
In other words this equation together with
(\ref{parttphi}) implies
\begin{equation}
I=\Phi \Psi'^2(\Phi(I)) \ ,
\end{equation}
where
$I=\frac{1}{4N^2}(\partial_t\phi)^2$.
Taking derivative this  equation with respect
to $I$ we find
\begin{equation}\label{Idrel}
1=\frac{d\Phi}{dI}\Psi'^2+
2\Phi\frac{d\Psi}{d\Phi}\frac{d^2\Psi}{d^2\Phi}
\frac{d\Phi}{dI}
\end{equation}
that will be useful below.

With the help of these results it is easy to
find the  Lagrangian
 in the form
\begin{eqnarray}
L&=&
\int d^D\bx (\partial_t\phi p+\partial_t g_{ij}
\pi^{ij}-N\mH_0-N^i\mH_i)=
\nonumber \\
&=& \int d^D\bx N\sqrt{g}\left(\frac
{1}{\kappa^2}K_{ij}
\mG^{ijkl}K_{kl}-\kappa^2 E^{ij}\mG_{ijkl}E^{kl}\right.+
\nonumber \\
&+& \left.\kappa^2 N\sqrt{g}\Psi(\Phi(I))-2\kappa^2 N\sqrt{g}
\frac{1}{\Psi'(\Phi(I))}I\right) \ , \nonumber \\
\end{eqnarray}
where $K_{ij}=\frac{1}{N}(\partial_t g_{ij}
-D_i N_j-D_j N_i)$.
Since we presumed that this Lagrangian
was  derived in the process of the gauge fixing
we easily find its general form when
  we perform the substitution
  \begin{equation}
  I=\frac{1}{4N^2}\partial_t\phi\partial_t\phi
\rightarrow \hat{I}=\frac{1}{4}(\nabla_n\phi)^2-
\Omega(g^{ij}\partial_i\phi\partial_j\phi) \ ,
\end{equation}
where
\begin{equation}
\nabla_n \phi=\frac{1}{N}(\partial_t \phi-N^i\partial_i\phi) \ ,
\end{equation} and where
$\Omega(x)$ is function that reflects the
anisotropy of the space-time
\cite{Capasso:2009fh,Suyama:2009vy,Romero:2009qs,Rama:2009xc}.
Then the  Lagrangian density takes the form
\begin{eqnarray}\label{Lgrsc}
\mL&=&N\sqrt{g}\left(\frac{1}{\kappa^2}K_{ij}
\mG^{ijkl}K_{kl}-\kappa^2 E^{ij}\mG_{ijkl}E^{kl}\right.+
\nonumber \\
&+&\kappa^2 \left.\Psi\left(\Phi\left(\hat{I}\right)\right)
-2\kappa^2 \frac{1}{\Psi'(\Phi(\hat{I}))}\hat{I}\right)
 \ . \nonumber \\
\end{eqnarray}
In order to check the consistency of our analysis we proceed in reverse
direction and determine Hamiltonian for $\phi$  from
(\ref{Lgrsc}).
The momentum $p_\phi$ conjugate to  $\phi$ takes the form
\begin{eqnarray}
p_\phi&=&
\frac{1}{2}\kappa^2 \sqrt{g}\nabla_n\phi(
 \Psi'\frac{d\Phi}{d \hat{I}}
+2\Psi''\Phi
\frac{d\Phi}{d\hat{I}}-
\frac{1}{\Psi'})=\nonumber \\
&=&-\frac{1}{2}\kappa^2 \sqrt{g}\nabla_n\phi
\frac{1}{\Psi'} \ ,
\nonumber \\
\end{eqnarray}
where $\Psi'=\frac{d\Psi}{d\Phi}$ and
where we used (\ref{Idrel}). Then the Hamiltonian
for $p_\phi$ takes the form
\begin{eqnarray}
H^{\phi}&=&
\int d^D\bx (N \mH^\phi_0+N^i\mH^\phi_i) \ ,
\nonumber \\
\mH_0^\phi
&=&-\kappa^2 \sqrt{g}\Psi\left(\frac{1}{\kappa^4 g}
p^2_\phi\right)-
\frac{1}{2}\kappa^2 \sqrt{g}\frac{1}{\Psi'(
\frac{1}{\kappa^2 g}p^2_\phi)}
\Omega\left(g^{ij}\partial_i\phi\partial_j\phi\right) \ , \quad
\mH^\phi_i=p_\phi\partial_i\phi \ .
 \nonumber \\
\end{eqnarray}
We see that this Hamiltonian constraint
$\mH_0^\phi$  coincides
with the $\phi-$part of the Hamiltonian
constraint (\ref{mH0}) (after fixing the gauge
$\phi=t$)  which justifies
our approach.

In summary, we  found the action of  Ho\v{r}ava-Lifshitz gravity
and scalar field that leads to RFDiff
invariant theory through the ghost condensation.
At this place we should stress one subtle point
in our analysis. It is well known that the
gauge fixing in the Hamiltonian framework
corresponds in the  imposing of the  additional constraints
(Gauge fixing functions)
on the system with the first class constraints
 such that
the Poisson brackets between gauge fixing functions
and the original first class constraints is non-zero
on the constraint surface. As a consequence
the gauge fixing functions together with the
original first class constraints become the second class constraints
that strongly vanish and can be explicitly solved
\footnote{For review, see
\cite{Henneaux:1992ig,Govaerts:2002fq,Govaerts:1991gd}.}.
In other words the gauge fixing makes  sense in case
when the constraint is the first class.  However
it was shown in
 \cite{Henneaux:2009zb} that the
 Hamiltonian constraint $\mH_0$ of the  Ho\v{r}ava-
 Lifshitz theory with space dependent lapse function
 is the second class constraint.
On the other hand  the projectable version of
Ho\v{r}ava-Lifshitz theory is characterized
the global form of the Hamiltonian constraint
\begin{equation}
\mathbf{H}=\int d^D \bx \mH_0 \ .
\end{equation}
Clearly $\pb{\mathbf{H},\mathbf{H}}=0$ and consequently
$\mathbf{H}$ can be considered as the first
class constraint.  Further the Poisson
bracket between $\mathbf{H}$ and $\mG(\bx)$ is  non-zero
and hence $\mathbf{H}$ together with
$\mG(\bx)$ form the second class constraints.
The the equation $\mathbf{H}=0$
can be solved with the  stronger
condition  $\mH_0(\bx)=0$
\footnote{For very nice analysis of this
issue, see \cite{Mukohyama:2009mz}.}
despite of the fact that these two
conditions are not equivalent in general.
In fact, the absence of the local form
of the Hamiltonian constraint in the
projectable version of Ho\v{r}ava-Lifshitz
theory has fatal consequence for the
spectrum of perturbative modes that
contain either tachyon or ghost modes
\cite{Weinfurtner:2010hz}.
It is clear that  the same problems occur
in the gauge fixed version of projectable
Ho\v{r}ava-Lifshitz gravity which is
RFDiff invariant Ho\v{r}ava-Lifshitz gravity
as was nicely shown in
\cite{Blas:2010hb}. However formally the
projectable Ho\v{r}ava-Lifshitz gravity is
well defined system with the Hamiltonian
given as a linear combination of the first
class constraints so that it is possible
to perform the gauge fixing that leads
to RFDiff invariant Ho\v{r}ava-Lifshitz
gravity.

.
\vskip 5mm

 \noindent {\bf
Acknowledgement:}
 This work   was also
supported by the Czech Ministry of
Education under Contract No. MSM
0021622409. \vskip 5mm


\begin{thebibliography}{20}
\bibitem{Horava:2009uw}
  P.~Horava,
\emph{``Quantum Gravity at a Lifshitz
Point,''}
  Phys.\ Rev.\  D {\bf 79} (2009) 084008
  [arXiv:0901.3775 [hep-th]].

\bibitem{Horava:2008jf}
  P.~Horava,
\emph{``Quantum Criticality and
Yang-Mills Gauge Theory,''}
  arXiv:0811.2217 [hep-th].


\bibitem{Horava:2008ih}
  P.~Horava,
\emph{``Membranes at Quantum
Criticality,''}
  JHEP {\bf 0903} (2009) 020
  [arXiv:0812.4287 [hep-th]].



\bibitem{Horava:2009if}
  P.~Horava,
\emph{``Spectral Dimension of the
Universe in Quantum Gravity at a
Lifshitz Point,''}
  Phys.\ Rev.\ Lett.\  {\bf 102} (2009) 161301
  [arXiv:0902.3657 [hep-th]].

\bibitem{Horava:2010zj}
  P.~Horava and C.~M.~Melby-Thompson,
\emph{``General Covariance in Quantum Gravity at a Lifshitz Point,''}
  arXiv:1007.2410 [hep-th].




\bibitem{Blas:2010hb}
  D.~Blas, O.~Pujolas and S.~Sibiryakov,
\emph{``Models of non-relativistic quantum gravity: the good, the bad and the
  healthy,''}
  arXiv:1007.3503 [hep-th].

\bibitem{Blas:2008uz}
  D.~Blas,
\emph{``Aspects of Infrared Modifications of Gravity,''}
  arXiv:0809.3744 [hep-th].

\bibitem{Bebronne:2009iy}
  M.~V.~Bebronne,
\emph{"Theoretical and phenomenological aspects of theories with massive
  gravitons,''}
  arXiv:0910.4066 [gr-qc].

\bibitem{Rubakov:2008nh}
  V.~A.~Rubakov and P.~G.~Tinyakov,
 \emph{``Infrared-modified gravities and massive gravitons,''}
  Phys.\ Usp.\  {\bf 51}, 759 (2008)
  [arXiv:0802.4379 [hep-th]].


\bibitem{Blas:2009yd}
  D.~Blas, O.~Pujolas and S.~Sibiryakov,
\emph{``On the Extra Mode and
 Inconsistency of Horava Gravity,''}
  JHEP {\bf 0910} (2009) 029
  [arXiv:0906.3046 [hep-th]].

\bibitem{Koyama:2009hc}
  K.~Koyama and F.~Arroja,
\emph{``Pathological
 behaviour of the scalar graviton in Ho\v{r}ava-Lifshitz
gravity,''}
  JHEP {\bf 1003} (2010) 061
  [arXiv:0910.1998 [hep-th]].

\bibitem{Cerioni:2010uz}
  A.~Cerioni and R.~H.~Brandenberger,
\emph{``Cosmological Perturbations in the Projectable Version of Horava-Lifshitz
  Gravity,''}
  arXiv:1007.1006 [hep-th].


\bibitem{ArkaniHamed:2003uy}
  N.~Arkani-Hamed, H.~C.~Cheng, M.~A.~Luty and S.~Mukohyama,
\emph{``Ghost condensation
and a consistent infrared modification of gravity,''}
  JHEP {\bf 0405} (2004) 074
  [arXiv:hep-th/0312099].


\bibitem{Blas:2009ck}
  D.~Blas, O.~Pujolas and S.~Sibiryakov,
\emph{``Comment on `Strong
 coupling in extended Horava-Lifshitz gravity',''}
  Phys.\ Lett.\  B {\bf 688} (2010) 350
  [arXiv:0912.0550 [hep-th]].

\bibitem{Blas:2009qj}
  D.~Blas, O.~Pujolas and S.~Sibiryakov,
\emph{``Consistent Extension Of Horava Gravity,''}
  Phys.\ Rev.\ Lett.\  {\bf 104} (2010) 181302
  [arXiv:0909.3525 [hep-th]].


\bibitem{Kluson:2010xx}
  J.~Kluson,
\emph{``Note About Hamiltonian
 Formalism of Modified $F(R)$ Ho\v{r}ava-Lifshitz
Gravities and Their Healthy Extension,''}
  Phys.\ Rev.\  D {\bf 82} (2010) 044004
  [arXiv:1002.4859 [hep-th]].


\bibitem{Kluson:2010nf}
  J.~Kluson,
\emph{``Note About Hamiltonian
 Formalism of Healthy Extended Horava-Lifshitz
Gravity,''}
  JHEP {\bf 1007} (2010) 038
  [arXiv:1004.3428 [hep-th]].






\bibitem{Kluson:2009rk}
  J.~Kluson,
\emph{``Horava-Lifshitz f(R) Gravity,''}
  JHEP {\bf 0911} (2009) 078
  [arXiv:0907.3566 [hep-th]].

\bibitem{Kluson:2009xx}
  J.~Kluson,
\emph{``New Models of f(R) Theories of Gravity,''}
  Phys.\ Rev.\  D {\bf 81} (2010) 064028
  [arXiv:0910.5852 [hep-th]].

\bibitem{Henneaux:2009zb}
  M.~Henneaux, A.~Kleinschmidt and G.~L.~Gomez,
\emph{``A dynamical inconsistency of Horava gravity,''}
  Phys.\ Rev.\  D {\bf 81} (2010) 064002
  [arXiv:0912.0399 [hep-th]].

\bibitem{Mukohyama:2009mz}
  S.~Mukohyama,
\emph{``Dark matter as
integration constant in Horava-Lifshitz gravity,''}
  Phys.\ Rev.\  D {\bf 80} (2009) 064005
  [arXiv:0905.3563 [hep-th]].



\bibitem{Capasso:2009fh}
  D.~Capasso and A.~P.~Polychronakos,
\emph{``Particle Kinematics in Horava-Lifshitz Gravity,''}
  JHEP {\bf 1002} (2010) 068
  [arXiv:0909.5405 [hep-th]].

\bibitem{Suyama:2009vy}
  T.~Suyama,
\emph{``Notes on Matter in Horava-Lifshitz Gravity,''}
  arXiv:0909.4833 [hep-th].

\bibitem{Romero:2009qs}
  J.~M.~Romero, V.~Cuesta, J.~A.~Garcia and J.~D.~Vergara,
\emph{``Conformal anisotropic
 mechanics and the Horava dispersion relation,''}
  Phys.\ Rev.\  D {\bf 81} (2010) 065013
  [arXiv:0909.3540 [hep-th]].

\bibitem{Rama:2009xc}
  S.~K.~Rama,
\emph{``Particle Motion
 with Ho\v{r}ava -- Lifshitz type Dispersion Relations,''}
  arXiv:0910.0411 [hep-th].

\bibitem{Henneaux:1992ig}
  M.~Henneaux and C.~Teitelboim,
\emph{``Quantization of gauge systems,''}
{\it  Princeton, USA: Univ. Pr. (1992) 520 p}


\bibitem{Govaerts:2002fq}
  J.~Govaerts,
\emph{``The quantum geometer's universe:
Particles, interactions and topology,''}
  arXiv:hep-th/0207276.

\bibitem{Govaerts:1991gd}
  J.~Govaerts,
\emph{``Hamiltonian
 Quantization And Constrained Dynamics,''}
{\it  Leuven, Belgium: Univ. Pr. (1991) 371 p. (Leuven notes in mathematical and theoretical physics, B4)}

\bibitem{Weinfurtner:2010hz}
  S.~Weinfurtner, T.~P.~Sotiriou and M.~Visser,
\emph{``Projectable Horava-Lifshitz gravity in a nutshell,''}
  J.\ Phys.\ Conf.\ Ser.\  {\bf 222} (2010) 012054
  [arXiv:1002.0308 [gr-qc]].



\bibitem{Chaichian:2010yi}
  M.~Chaichian, S.~Nojiri, S.~D.~Odintsov, M.~Oksanen and A.~Tureanu,
\emph{``Modified F(R)
Horava-Lifshitz gravity: a way to accelerating FRW
cosmology,''}
  Class.\ Quant.\ Grav.\  {\bf 27} (2010) 185021
  [arXiv:1001.4102 [hep-th]].

\bibitem{Carloni:2010nx}
  S.~Carloni, M.~Chaichian, S.~Nojiri, S.~D.~Odintsov, M.~Oksanen and A.~Tureanu,
\emph{``Modified first-order Horava-Lifshitz gravity:
Hamiltonian analysis of the
  general theory and accelerating FRW cosmology in power-law F(R) model,''}
  arXiv:1003.3925 [hep-th].

\bibitem{Chaichian:2010zn}
  M.~Chaichian, M.~Oksanen and A.~Tureanu,
\emph{``Hamiltonian analysis
 of non-projectable modified F(R) Ho\v{r}ava-Lifshitz
gravity,''}
  arXiv:1006.3235 [hep-th].

\bibitem{Elizalde:2010ep}
  E.~Elizalde, S.~Nojiri, S.~D.~Odintsov and D.~Saez-Gomez,
\emph{``Unifying inflation with dark energy in modified F(R) Horava-Lifshitz
  gravity,''}
  arXiv:1006.3387 [hep-th].











\end{thebibliography}
\end{document}